\documentclass[11pt]{article}
\usepackage[a4paper,margin=24mm]{geometry}
\usepackage[T1]{fontenc}
\usepackage{lmodern,microtype,amsmath,amssymb,bm,graphicx,booktabs,tabularx}
\usepackage[numbers,sort&compress]{natbib}
\usepackage[dvipsnames]{xcolor}
\usepackage{listings}
\usepackage[colorlinks=true,linkcolor=MidnightBlue,citecolor=MidnightBlue,urlcolor=MidnightBlue]{hyperref}
\usepackage{placeins}

\newcommand{\pkg}{\textsc{MagBerry}}
\newcommand{\dd}{\mathrm d}
\newcommand{\kb}{k_{\mathrm B}}
\newcommand{\BZ}{\mathrm{BZ}}
\newcommand{\figfile}[2]{\includegraphics[width=#2\linewidth]{#1.pdf}}
\newcommand{\PAMMesh}{10}

\newcommand{\PAMHundred}{0.0953}
\newcommand{\PAMHundredFifty}{0.104}

\title{MagBerry: A Julia package for modeling magnon and phonon topology and transport}
\author{Xiaonong Shen and Wei Ren\thanks{Corresponding author: \href{mailto:renwei@shu.edu.cn}{renwei@shu.edu.cn}.}\\[0.4em]
\small Physics Department, Materials Genome Institute,\\
\small State Key Laboratory of Advanced Refractories,\\
\small Institute for Quantum Science and Technology,\\
\small Shanghai University, Shanghai 200444, China}
\date{25 October 2026}
\begin{document}
\maketitle
\begin{abstract}
We present \pkg, a Julia package for studying the topology and transport of magnons, phonons, and their hybrid excitations. It combines magnetic models and linear spin-wave Hamiltonians from Sunny.jl with native modules for harmonic phonons and magnon--phonon coupling. These excitations are treated within a common bosonic Bogoliubov--de Gennes formulation. The package calculates bulk and boundary spectra, Berry curvature, and Chern numbers, and resolves the spin and phonon angular momentum carried by each mode. Transport calculations include thermal Hall, spin and orbital Nernst, and phonon angular momentum Hall responses. The package also evaluates the extended Berry curvature dipole contribution to nonlinear magnon thermal Hall transport. Relaxation-time Boltzmann theory gives heat and angular momentum currents. Examples based on published models reproduce topological magnon boundary spectra and illustrate phonon geometry and nonlinear transport. For an altermagnetic hybrid model, we extend a published study of the $d$-wave phonon angular momentum texture by calculating its Boltzmann transport coefficients. The resulting tensor describes how the longitudinal and transverse angular momentum currents vary with temperature and the direction of the temperature gradient.
\end{abstract}
\section*{Program summary}
\begin{description}
\item[Program title:] MagBerry.
\item[Programming language:] Julia; the reported calculations use Julia 1.11.3.
\item[External libraries:] Sunny 0.9.2, CairoMakie 0.15.13, and dependencies specified by the accompanying environment.
\item[Version used:] MagBerry 0.2.4.
\item[Licensing provisions:] MIT.
\item[Nature of problem:] Calculation of excitation spectra, geometric observables, and transport coefficients in quadratic models of magnons, phonons, and their coupled modes.
\item[Solution method:] Bosonic diagonalization with the commutation metric, Kubo and lattice-link calculations of Berry curvature, generalized-current response, relaxation-time Boltzmann transport, and direct diagonalization of finite geometries.
\item[Restrictions:] Harmonic phonons and linear spin-wave theory about a specified reference state. The generic boundary interface retains bulk onsite blocks. Free surfaces require explicit reconstruction and a reference-state check.
\item[Developer's repository:] \url{https://github.com/Raito-SHU/MagBerry.jl}.
\item[Documentation:] \url{http://renwei.group/softwares/MagBerry.jl/en/}.
\end{description}
\section{Introduction}
Magnons and phonons are electrically neutral bosonic excitations of spins and the lattice, respectively. They carry energy and angular momentum without an accompanying charge current, enabling information transport in insulators. Their bands can also have nontrivial topology. Neutron scattering has provided evidence for topological magnon bands in Cu(1,3-bdc) and CrI$_3$~\cite{chisnell2015,chen2018}. For phonons, inelastic X-ray scattering combined with theoretical calculations has identified double Weyl nodes in FeSi~\cite{miao2018}. Both excitations can exhibit transverse thermal transport analogous to the electronic Hall effect. Strohm et~al. discovered the phonon Hall effect in Tb$_3$Ga$_5$O$_{12}$ in 2005~\cite{strohm2005}, and Onose et~al. observed the magnon thermal Hall effect in the ferromagnetic insulator Lu$_2$V$_2$O$_7$ in 2010~\cite{onose2010}. Theory has also predicted transverse angular momentum transport driven by temperature gradients, including the magnon spin Nernst effect in collinear antiferromagnets~\cite{cheng2016} and the phonon angular momentum Hall effect in lattice systems~\cite{parkyang}.

Magnon--phonon coupling can produce hybrid excitations known as magnon polarons, which combine magnetic and lattice motion. In FePS$_3$, magneto-Raman measurements have revealed strong coupling~\cite{liu2021feps3}, and terahertz pulses have enabled coherent excitation of magnon polarons~\cite{luo2025feps3}. Spectroscopy of FePSe$_3$ has identified chirality-selective hybridization and chiral magnon polarons at zero magnetic field~\cite{cui2023fepse3}. Neutron scattering combined with theoretical analysis has also identified topological magnon polarons in Fe$_2$Mo$_3$O$_8$~\cite{bao}. In $d$-wave altermagnets, theory predicts that the energy splitting between magnon branches of opposite chirality changes sign with momentum. Chirality-selective hybridization allows the phonon components of the hybrid modes to acquire a corresponding $d$-wave angular momentum texture~\cite{bendin}.

Studies of chiral phonons and altermagnetic magnons have increased the demand for calculations of band topology and angular momentum transport. SpinW and Sunny.jl are widely used to calculate magnetic excitation spectra and fit experimental data. SpinW computes linear spin-wave spectra from spin models~\cite{spinw2015}, while Sunny.jl supports multiflavor spin-wave theory and dynamical simulations~\cite{sunny}. For phonons, PH-NODE searches for topological band nodes using first-principles results~\cite{phnode2024}. Calculating group velocities, Berry curvature, boundary states, and transport responses such as the thermal and spin Hall effects from these models generally requires additional post-processing. Publicly available tools for both topology and transport in coupled magnon--phonon systems remain limited.

We developed \pkg, a Julia package for calculating the topology and transport of magnons, phonons, and magnon polarons. The package treats all three types of excitation through quadratic bosonic Bogoliubov--de Gennes (BdG) Hamiltonians~\cite{colpa,matsumoto2014}. Sunny.jl supplies the magnetic models and their linear spin-wave Hamiltonians. Native \pkg\ modules construct harmonic phonon and magnon--phonon Hamiltonians and solve for the hybrid modes. Structures, force constants, and magnetic interaction parameters can be imported from first-principles outputs. The magnon routines calculate band dispersions and group velocities and resolve spin polarization, chirality, and wave-packet orbital moments. They also evaluate Berry curvature, Chern numbers, and boundary states. Bulk and boundary spectra and Berry curvature are available for phonon and hybrid models as well. For hybrid modes, \pkg\ separates the magnetic and lattice components and resolves their spin and phonon angular momentum.

The transport routines evaluate linear thermal Hall, spin Nernst, magnon orbital Nernst, and phonon angular momentum Hall coefficients. They also compute the extended Berry curvature dipole contribution to nonlinear magnon thermal Hall transport. Within the relaxation-time approximation, Boltzmann calculations give heat, spin, and phonon angular momentum currents as functions of temperature and the direction of the temperature gradient. For hybrid modes, the spin and phonon angular momentum contributions to transport can be resolved separately.

\section{Models and computational methods}
\label{sec:methods}
Throughout this section, $n,m$ label modes, $\mu,\nu$ label atoms, and $i,j,a$ label Cartesian components. In a transport tensor, $i$ specifies the current direction, $j$ the temperature-gradient direction, and $a$ the polarization of the transported observable. We set $\hbar=1$ in the formulas below and restore physical units in the numerical outputs.

\subsection{Bosonic BdG Hamiltonians and eigenmodes}
\label{sec:hamiltonians}
Magnon, phonon, and hybrid excitations are obtained by expanding the spin and lattice Hamiltonian to quadratic order about a reference configuration. The inputs are the magnetic interactions, reference spin configuration, atomic masses, force constants, and magnon--phonon coupling parameters. Additive constants are omitted from the Hamiltonians.

Sunny.jl constructs the magnetic model and its quadratic spin-wave Hamiltonian~\cite{sunny,holstein1940}. Let $\bm a_{\bm k}$ be the column vector of $N_{\mathrm m}$ magnon annihilation operators. In the Nambu basis $\Psi_{\mathrm m,\bm k}=(\bm a_{\bm k}^{T},\bm a_{-\bm k}^{\dagger})^T$, the Hamiltonian is $\hat H_{\mathrm m}^{(2)}=\tfrac12\sum_{\bm k}\Psi_{\mathrm m,\bm k}^{\dagger}\mathcal H_{\mathrm m}(\bm k)\Psi_{\mathrm m,\bm k}$. Its matrix has the block form
\begin{equation}
\mathcal H_{\mathrm m}(\bm k)=
\begin{pmatrix}A(\bm k)&B(\bm k)\\B^{\dagger}(\bm k)&A^T(-\bm k)\end{pmatrix}.
\label{eq:magnetic_bdg}
\end{equation}
The blocks $A=A^{\dagger}$ and $B(\bm k)=B^T(-\bm k)$ collect number-conserving and pairing terms, respectively. \pkg\ directly uses the Hermitian matrix $\mathcal H_{\mathrm m}$ supplied by Sunny. The Nambu commutator matrix is $\mathcal G_{\mathrm m}=\operatorname{diag}(I,-I)$.

The phonon model uses atomic masses $M_\mu$ and harmonic force constants $\Phi_{\mu i,\nu j}(\bm R)$. These force constants couple displacement component $i$ of atom $\mu$ in the reference cell to component $j$ of atom $\nu$ in cell $\bm R$. Fourier transformation gives the mass-weighted dynamical matrix~\cite{phonopy2023}
\begin{equation}
D^{(\omega)}_{\mu i,\nu j}(\bm k)=\frac{1}{\sqrt{M_\mu M_\nu}}
\sum_{\bm R}\Phi_{\mu i,\nu j}(\bm R)e^{i\bm k\cdot\bm R}.
\label{eq:phonon_dynamical}
\end{equation}
The canonical solver uses the squared-energy matrix $D^{(E)}=\hbar^2D^{(\omega)}$, denoted by $D$ below. With $\hbar=1$, introduce mass-weighted coordinates $Q_{\mu i}=\sqrt{M_\mu}u_{\mu i}$ and $P_{\mu i}=p_{\mu i}/\sqrt{M_\mu}$. For $N_{\mathrm{ph}}$ retained displacement coordinates per cell, define $X_{\bm k}=(\bm Q_{\bm k}^T,\bm P_{\bm k}^T)^T$. The harmonic Hamiltonian $\hat H_{\mathrm{ph}}=\tfrac12\sum_{\bm k}X_{\bm k}^{\dagger}\mathcal H_{\mathrm{ph}}(\bm k)X_{\bm k}$ has the matrix
\begin{equation}
\mathcal H_{\mathrm{ph}}(\bm k)=\begin{pmatrix}D(\bm k)&0\\0&I\end{pmatrix}.
\label{eq:phonon_hamiltonian}
\end{equation}
The commutator matrix is $\mathcal G_{\mathrm{ph}}=\left(\begin{smallmatrix}0&iI\\-iI&0\end{smallmatrix}\right)$. This coordinate--momentum representation is equivalent to a bosonic creation--annihilation basis.

Magnon--phonon coupling follows from the displacement dependence of magnetic interactions. Expanding the exchange tensor to first order in displacement gives
\begin{equation}
\mathsf J_{\mu\nu}(\{\bm u\})=\mathsf J_{\mu\nu}^{(0)}+
\sum_{\lambda a}\frac{\partial\mathsf J_{\mu\nu}}{\partial u_{\lambda a}}u_{\lambda a}+\cdots ,
\label{eq:exchange_displacement}
\end{equation}
where $\lambda$ labels a displaced atom. Terms linear in both displacement and spin fluctuation form the bilinear coupling matrix. The tensor derivatives include symmetric exchange and Dzyaloshinskii--Moriya contributions. Alternatively, an explicit magnetoelastic model can supply the coupling matrix~\cite{go}. In the combined basis $\Xi_{\bm k}=(\Psi_{\mathrm m,\bm k}^T,X_{\bm k}^T)^T$, the quadratic Hamiltonian is $\hat H_2=\tfrac12\sum_{\bm k}\Xi_{\bm k}^{\dagger}\mathcal H(\bm k)\Xi_{\bm k}$, with
\begin{equation}
\mathcal H(\bm k)=
\begin{pmatrix}\mathcal H_{\mathrm m}(\bm k)&C(\bm k)\\
C^{\dagger}(\bm k)&\mathcal H_{\mathrm{ph}}(\bm k)\end{pmatrix}.
\label{eq:hybrid_hamiltonian}
\end{equation}
Here $C=(C_Q,C_P)$ follows the displacement--momentum ordering of $X_{\bm k}$. Couplings obtained from displacement derivatives enter $C_Q$. Setting $C=0$ recovers the uncoupled models.

The canonical commutator is $\mathcal G=\operatorname{diag}(\mathcal G_{\mathrm m},\mathcal G_{\mathrm{ph}})$. The excitation energies and mode vectors satisfy~\cite{colpa}
\begin{equation}
\mathcal G\mathcal H(\bm k)v_{n\bm k}=E_{n\bm k}v_{n\bm k},\qquad
v_{n\bm k}^{\dagger}\mathcal Gv_{m\bm k}=s_n\delta_{nm},\quad s_n=\pm1.
\label{eq:modes}
\end{equation}
For a stable reference configuration, finite positive-energy modes have $s_n=+1$. The notation $n>0$ below selects these branches. Collecting the complete set of finite-frequency mode vectors as $U=(v_1,v_2,\ldots)$ gives $U^{\dagger}\mathcal GU=\eta=\operatorname{diag}(s_n)$, the commutator in the mode basis.

\subsection{Berry curvature, topology, and extended dipoles}
\label{sec:geometry}
For an isolated positive-energy band, the Berry connection is $\mathcal A_{n,i}=-i v_{n\bm k}^{\dagger}\mathcal G\partial_{k_i}v_{n\bm k}$, and $\Omega_{n,ij}=\partial_{k_i}\mathcal A_{n,j}-\partial_{k_j}\mathcal A_{n,i}$ is its ordinary Berry curvature~\cite{berry1984,shindou2013}. Cartesian momentum derivatives include the phase factors associated with the atomic basis positions. In the mode basis, define the velocity matrix $V_i=U^{\dagger}(\partial_{k_i}\mathcal H)U$. The Kubo expression is
\begin{equation}
\Omega_{n,ij}=2\operatorname{Im}\sum_{m\ne n}
s_ns_m\frac{(V_i)_{nm}(V_j)_{mn}}{(E_n-E_m)^2}.
\label{eq:ordinary_curvature}
\end{equation}
All energies and matrix elements in the Kubo expressions are evaluated at the same wave vector. The intermediate-state sum includes modes in both the positive- and negative-frequency sectors. For an isolated band over a two-dimensional Brillouin zone, the Chern number is~\cite{tknn1982}
\begin{equation}
C_n=\frac{1}{2\pi}\int_{\BZ}\Omega_{n,xy}(\bm k)\,\dd^2k.
\label{eq:chern}
\end{equation}
\pkg\ evaluates ordinary curvature and Chern numbers using either Eq.~\eqref{eq:ordinary_curvature} or the discrete lattice-link method of Fukui, Hatsugai, and Suzuki~\cite{fukui}.

To describe spin and angular momentum transport, let $O$ denote a selected Cartesian component, such as $S_a$ or $L_{\mathrm{ph},a}$. The laboratory-frame spin operator is assembled from the local spin coordinates. Mechanical phonon angular momentum (PAM) is $\hat{\bm L}_{\mathrm{ph}}=\sum_\mu\hat{\bm u}_\mu\times\hat{\bm p}_\mu$~\cite{zhangniu2014}. If $\mathcal O$ is the matrix of the quadratic observable in the canonical basis, its mode-basis matrix is $O=U^{\dagger}\mathcal O U$. For $O\in\{S_a,L_{\mathrm{ph},a}\}$, the symmetrized current matrix is $J_i^O=\tfrac12(V_i\eta O+O\eta V_i)$. Its interband matrix elements give the current Berry curvature
\begin{equation}
\Omega_{n,ij}^{O}=2\operatorname{Im}\sum_{m\ne n}
s_ns_m\frac{(J_i^{O})_{nm}(V_j)_{mn}}{(E_n-E_m)^2}.
\label{eq:current_curvature}
\end{equation}
The polarization index is $a$, the current direction is $i$, and the driving direction is $j$. The interband matrix elements in these curvatures determine the spin Nernst and phonon angular momentum Hall responses~\cite{zyuzin2016,parkyang}.

For orbital transport, $O=M_{\mathrm{orb},a}$ is a component of the magnon wave-packet moment $\hat{\bm M}_{\mathrm{orb}}=\tfrac14(\hat{\bm r}\times\hat{\bm v}-\hat{\bm v}\times\hat{\bm r})$. Its mode-basis matrix is constructed from velocity matrix elements. We adopt the orbital-current matrix $J_i^{M_{\mathrm{orb},a}}=\tfrac14(V_i\eta M_{\mathrm{orb},a}+M_{\mathrm{orb},a}\eta V_i)$ and the curvature convention of Go et~al.~\cite{goorbital2024}. The orbital-current Berry curvature is then
\begin{equation}
\Omega_{n,ij}^{M_{\mathrm{orb},a}}=-2\operatorname{Im}\sum_{m\ne n}
s_ns_m\frac{(J_i^{M_{\mathrm{orb},a}})_{nm}(V_j)_{mn}}{(E_n-E_m)^2}.
\label{eq:orbital_curvature}
\end{equation}
The corresponding thermal prefactors are specified in Sec.~\ref{sec:transport}. At degeneracies, the geometric quantities are evaluated through subspace traces~\cite{wilczek1984}.

The extended-dipole contribution to nonlinear magnon transport depends on the momentum derivative of the energy-weighted Berry curvature. With $\Omega_n^b=\tfrac12\sum_{ij}\epsilon_{bij}\Omega_{n,ij}$, the extended Berry curvature dipole field is~\cite{kondo2022}
\begin{equation}
\overline D_{n,i}^{\,b}=\partial_{k_i}\!\left(E_n\Omega_n^b\right)
=(\partial_{k_i}E_n)\Omega_n^b+E_n\partial_{k_i}\Omega_n^b.
\label{eq:extended_bcd}
\end{equation}
Here $i$ labels the derivative direction and $b$ the curvature component. \pkg\ evaluates this field on a periodic momentum grid and can weight it by $E_n$ or the mode spin expectation $\langle S_a\rangle_n$. The energy-weighted field $E_n\overline D_{n,i}^{\,b}$ enters the nonlinear thermal Hall coefficient~\cite{ni2025}.

\subsection{Boltzmann and Hall transport coefficients}
\label{sec:transport}
\begin{subequations}
\label{eq:linear_response}
The thermal conductivity tensor relates the heat current to the temperature gradient,
\begin{equation}
j_i^E=-\sum_j\kappa_{ij}(T)\,\partial_jT.
\end{equation}

For a selected observable $O$, including its polarization as defined in Sec.~\ref{sec:geometry}, we define the corresponding response tensor by
\begin{equation}
j_i^O=-\sum_j\Lambda_{ij}^{O}(T)\,\partial_jT.
\end{equation}
\end{subequations}

Superscripts $\mathrm B$ and $\mathrm H$ denote Boltzmann and intrinsic Hall contributions, respectively. For a fixed Hamiltonian, the Bose occupation is $\rho_{n\bm k}(T)=n_B(E_{n\bm k},T)=[\exp(E_{n\bm k}/(\kb T))-1]^{-1}$. Brillouin-zone integrals use the measure $\int_{\bm k}=\int_{\BZ}\dd^dk/(2\pi)^d$, where $d$ is the spatial dimension.

In the relaxation-time approximation, a temperature gradient produces the occupation correction $\delta\rho_{n\bm k}=-\tau\sum_jv_{n,j}^{\mathrm g}(\partial_T\rho_{n\bm k})\partial_jT$~\cite{flebus2017}. Here $\tau$ is the common relaxation time and $v_{n,i}^{\mathrm g}=\partial_{k_i}E_{n\bm k}=(V_i)_{nn}$ is the group velocity of a positive-norm mode.

\begin{subequations}
Each mode carries an energy flux $E_{n\bm k}v_{n,i}^{\mathrm g}$, giving the Boltzmann thermal conductivity
\begin{equation}
\kappa_{ij}^{\mathrm B}(T)=\tau\sum_{n>0}\int_{\bm k}
E_{n\bm k}v_{n,i}^{\mathrm g}v_{n,j}^{\mathrm g}\,\partial_T\rho_{n\bm k}.
\end{equation}

For spin or phonon angular momentum, $O\in\{S_a,L_{\mathrm{ph},a}\}$, the mode current is $\mathcal J_{n,i}^{O}=(J_i^O)_{nn}$, with the current matrix defined before Eq.~\eqref{eq:current_curvature}. The resulting response coefficient is
\begin{equation}
\Lambda_{ij}^{O,\mathrm B}(T)=\tau\sum_{n>0}\int_{\bm k}
\mathcal J_{n,i}^{O}v_{n,j}^{\mathrm g}\,\partial_T\rho_{n\bm k}.
\label{eq:boltzmann}
\end{equation}
\end{subequations}

The choice of observable resolves the spin and phonon angular momentum contributions of a hybrid mode. Within a degenerate subspace, operator traces replace the single-band products.

Intrinsic Hall responses use the thermal functions $c_1(\rho)=(1+\rho)\ln(1+\rho)-\rho\ln\rho$ and $c_2(\rho)=\int_0^\rho[\ln(1+t^{-1})]^2\dd t$. The ordinary curvature gives the thermal Hall coefficient~\cite{matsumoto2011,matsumoto2011b,matsumoto2014},
\begin{equation}
\kappa_{ij}^{\mathrm H}(T)=-\kb^2T\sum_{n>0}\int_{\bm k}
c_2(\rho_{n\bm k})\,\Omega_{n,ij}(\bm k).
\label{eq:thermal_hall}
\end{equation}

\begin{subequations}
\label{eq:intrinsic_observable_transport}
The spin Nernst coefficient describes the transverse flow of spin component $S_a$ driven by a temperature gradient~\cite{zyuzin2016}. In terms of the spin-current curvature from Eq.~\eqref{eq:current_curvature}, the coefficient is
\begin{equation}
\Lambda_{ij}^{S_a,\mathrm H}(T)
=-\kb\sum_{n>0}\int_{\bm k}c_1(\rho_{n\bm k})\,\Omega_{n,ij}^{S_a}(\bm k).
\label{eq:spin_nernst}
\end{equation}

The phonon angular momentum Hall response is determined by the current curvature associated with $L_{\mathrm{ph},a}$~\cite{parkyang}. Its coefficient is
\begin{equation}
\Lambda_{ij}^{L_{\mathrm{ph},a},\mathrm H}(T)
=\kb\sum_{n>0}\int_{\bm k}c_1(\rho_{n\bm k})\,\Omega_{n,ij}^{L_{\mathrm{ph},a}}(\bm k).
\label{eq:beta}
\end{equation}

The magnon orbital Nernst response transports the wave-packet moment $M_{\mathrm{orb},a}$. With the orbital-current normalization and curvature in Eq.~\eqref{eq:orbital_curvature}, its coefficient reads~\cite{goorbital2024}
\begin{equation}
\Lambda_{ij}^{M_{\mathrm{orb},a},\mathrm H}(T)
=2\kb\sum_{n>0}\int_{\bm k}c_1(\rho_{n\bm k})\,\Omega_{n,ij}^{M_{\mathrm{orb},a}}(\bm k).
\label{eq:orbital_nernst}
\end{equation}
\end{subequations}

In the examples, we write $\beta_{ij}^{a}\equiv\Lambda_{ij}^{L_{\mathrm{ph},a},\mathrm H}$ for the intrinsic phonon response and $\beta_{ij}^{L_a}\equiv\Lambda_{ij}^{L_{\mathrm{ph},a},\mathrm B}$ for its Boltzmann contribution. Temperature scans reuse the calculated energies and current curvatures, updating only the Bose weights.

For the two-dimensional nonlinear magnon thermal Hall response, we define $j_x^{E,(2)}=\chi_{xyy}(T)(\partial_yT)^2$. The extended-dipole contribution in the relaxation-time approximation is~\cite{ni2025}
\begin{equation}
\chi_{xyy}(T)=\frac{2\tau\kb}{T}\sum_{n>0}\int_{\bm k}
c_1(\rho_{n\bm k})\,E_{n\bm k}\overline D_{n,y}^{\,z}(\bm k),
\label{eq:nonlinear_thc}
\end{equation}
where $\overline D_{n,y}^{\,z}$ is defined by Eq.~\eqref{eq:extended_bcd}.

\subsection{Boundary spectra and localization}
\label{sec:boundary_method}
We construct the Hamiltonian of a finite geometry by partially Fourier transforming the bulk BdG matrix. Strips retain one periodic direction and a finite width, while slabs retain two periodic directions and a finite thickness. \pkg\ applies this construction to magnons, phonons, and their hybrid modes.

For one finite direction, denote reciprocal-lattice coordinates by $(\bm q_{\parallel},q_{\perp})$. In the cell-periodic basis, the intercell matrix at integer separation $r$ is
\begin{equation}
h_r(\bm q_{\parallel})=\int_0^1\dd q_{\perp}\,
e^{-2\pi i q_{\perp}r}\mathcal H(\bm q_{\parallel},q_{\perp}).
\label{eq:boundary_fourier}
\end{equation}
The integral is evaluated on a discrete grid. Retaining $W$ cells gives $[\mathcal H_W]_{\ell\ell'}=h_{\ell'-\ell}$ for $\ell,\ell'=1,\ldots,W$. Couplings across the sample edge are truncated, and the default termination retains the bulk onsite blocks. In cellwise basis ordering, $\mathcal G_W=I_W\otimes\mathcal G$. The finite-system modes follow from $\mathcal G_W\mathcal H_Wv_n=E_nv_n$ and are normalized by $v_n^{\dagger}\mathcal G_Wv_n=1$ for positive-energy modes.

To characterize localization, let $v_{n\ell}$ be the part of a mode vector in cell $\ell$. A fixed local oscillator basis defines the positive weight matrix $\mathsf W_{\mathrm{loc}}=\operatorname{diag}(I,I,\boldsymbol\omega_0,\boldsymbol\omega_0^{-1})$ in the $(a,a^{\dagger},Q,P)$ ordering, where $\boldsymbol\omega_0$ is a diagonal matrix of positive reference frequencies. Only the blocks corresponding to sectors present in the model are retained. The normalized cell weight is
\begin{equation}
w_{n\ell}=\frac{v_{n\ell}^{\dagger}\mathsf W_{\mathrm{loc}}v_{n\ell}}
{\sum_{\ell'}v_{n\ell'}^{\dagger}\mathsf W_{\mathrm{loc}}v_{n\ell'}},
\qquad \sum_\ell w_{n\ell}=1.
\label{eq:boundary_weight}
\end{equation}
For an edge or surface region $\mathcal E$ comprising selected outermost cells, $w_{n,\mathcal E}=\sum_{\ell\in\mathcal E}w_{n\ell}$. Its localization-weighted spectral intensity is
\begin{equation}
A_{\mathcal E}(\bm q_{\parallel},E)=\sum_{n>0}w_{n,\mathcal E}(\bm q_{\parallel})
\frac{\gamma/\pi}{[E-E_n(\bm q_{\parallel})]^2+\gamma^2},
\label{eq:boundary_spectrum}
\end{equation}
where $\gamma>0$ is the energy broadening. The boundary dispersion and weights, together with bulk gaps and Chern numbers, characterize bulk--boundary correspondence~\cite{shindou2013}.

Transforming two directions constructs finite two-dimensional samples or rods periodic along the third direction. Their eigenmodes resolve localization at corners or hinges. For hybrid modes, \pkg\ also evaluates magnetic and lattice composition. The observable matrix elements defined in Sec.~\ref{sec:geometry} give the spatial distributions of spin and phonon angular momentum.

\section{Software organization and use}
\label{sec:software}
\subsection{Installation and runtime environment}
\pkg\ requires Julia 1.10 or later. Extract the source package and run the following commands in Julia, using the source directory that contains \texttt{Project.toml}.
\begin{lstlisting}
using Pkg
Pkg.activate()
Pkg.develop(path="path/to/MagBerry.jl")
Pkg.add(["Sunny", "CairoMakie"])
\end{lstlisting}
Replace \texttt{path/to/MagBerry.jl} with that directory. The package manager installs the dependencies. To begin a calculation script, load the packages:
\begin{lstlisting}
using Sunny, MagBerry, CairoMakie
\end{lstlisting}
Each input script constructs a model, specifies momentum sampling and calculation parameters, and calls the analysis functions.

\subsection{Model construction and data input}
Three model types represent magnons, phonons, and hybrid modes. Sunny constructs magnetic models and provides their spin-wave Hamiltonians through \texttt{SpinWaveTheory}~\cite{sunny}. Phonon models use a crystal structure, atomic masses, and force constants; hybrid models combine the magnetic and phonon models with their coupling. Table~\ref{tab:model_input} lists the constructors and input interfaces. To map supercell force constants onto the model cell, the importer uses the structures of both cells~\cite{phonopy2023}.

\begin{table}[!htbp]
\centering
\caption{Model construction and data input interfaces.}
\label{tab:model_input}
\small
\renewcommand{\arraystretch}{1.12}
\setlength{\tabcolsep}{5pt}
\begin{tabularx}{\linewidth}{@{}>{\raggedright\arraybackslash}p{0.46\linewidth}>{\raggedright\arraybackslash}X@{}}
\toprule
Function or constructor & Description \\
\midrule
\texttt{Sunny.SpinWaveTheory} & Construct a spin-wave model from magnetic interactions and a reference spin configuration. \\
\texttt{PhononSystem} & Create a phonon model from a crystal structure and atomic masses. \\
\texttt{HybridSystem} & Combine spin waves, phonons, and magnon--phonon coupling. \\
\texttt{set\_force!}, \texttt{set\_spring!} & Set force-constant matrices or spring interactions. \\
\texttt{read\_poscar\_crystal} & Read a crystal structure from a POSCAR file. \\
\texttt{read\_phonopy\_phonon} & Build a phonon model from crystal and supercell structures and Phonopy force constants. \\
\bottomrule
\end{tabularx}
\end{table}

\subsection{Calculation interfaces and workflow}
Each calculation takes a model, a momentum path or Brillouin-zone grid, and the selected bands or response components. Spectra, boundary modes, and Boltzmann transport are calculated directly from the model. Hall and Nernst calculations integrate the corresponding curvatures over the Brillouin zone, applying Bose weights at each requested temperature. Table~\ref{tab:analysis_functions} lists the main functions.

\begin{table}[!htbp]
\centering
\caption{Main analysis and transport functions.}
\label{tab:analysis_functions}
\small
\renewcommand{\arraystretch}{1.12}
\setlength{\tabcolsep}{5pt}
\begin{tabularx}{\linewidth}{@{}>{\raggedright\arraybackslash}p{0.46\linewidth}>{\raggedright\arraybackslash}X@{}}
\toprule
Function & Description \\
\midrule
\texttt{Sunny.dispersion}\newline\texttt{phonon\_bands}\newline\texttt{hybrid\_bands} & Calculate bulk magnon, phonon, and hybrid spectra, respectively. \\
\texttt{berry\_curvature} & Calculate ordinary Berry curvature or the curvature associated with spin, orbital moment, and phonon angular momentum currents. \\
\texttt{chern\_numbers} & Calculate band Chern numbers for magnon, phonon, and hybrid models. \\
\texttt{phonon\_angular\_momentum}\newline\texttt{hybrid\_mode\_observables} & Evaluate phonon angular momentum and the composition and angular momenta of hybrid modes. \\
\texttt{boltzmann\_transport} & Calculate heat, spin, and phonon angular momentum currents driven by a temperature gradient. \\
\texttt{thermal\_hall\_conductivity} & Calculate the linear thermal Hall conductivity. \\
\texttt{spin\_nernst\_conductivity}\newline\texttt{orbital\_nernst\_conductivity} & Calculate spin Nernst and magnon orbital Nernst coefficients, respectively. \\
\texttt{angular\_momentum\_\allowbreak nernst\_conductivity} & Calculate the phonon angular momentum Hall coefficient. \\
\texttt{magnon\_extended\_bcd} & Construct the extended magnon Berry curvature dipole field. \\
\texttt{magnon\_nonlinear\_\allowbreak thermal\_hall\_\allowbreak conductivity} & Calculate the extended-dipole contribution to nonlinear magnon thermal Hall transport. \\
\texttt{boundary\_bands} & Calculate boundary spectra for strips, slabs, finite two-dimensional samples, and rods. \\
\bottomrule
\end{tabularx}
\end{table}

The \texttt{chern\_numbers} interface accepts a model or previously calculated ordinary Berry-curvature data. The \texttt{boundary\_bands} interface selects the geometry through \texttt{geometry}, with values \texttt{:strip}, \texttt{:surface}, \texttt{:finite2d}, and \texttt{:hinge}. For phonon and hybrid models, the phonon angular momentum Hall interface offers native adaptive integration over a requested temperature range. Appendix~\ref{app:example_input} gives a complete phonon input. Scripts for the four main examples are provided in \texttt{examples/paper}; further tutorials are available in the \href{http://renwei.group/softwares/MagBerry.jl/en/}{online documentation}.
\FloatBarrier

OpenAI Codex (OpenAI, used in October 2026) assisted development of the adaptive integration and boundary interfaces and preparation of example and plotting scripts. The numerical results and figures were generated by the Julia implementations and inputs supplied with the program.

\section{Illustrative examples}
Four examples based on published models demonstrate magnon topology and boundary states, linear and nonlinear thermal Hall responses, and phonon angular momentum transport in elastic and hybrid systems. The \href{http://renwei.group/softwares/MagBerry.jl/en/}{online documentation} provides complete inputs and further examples, including spin Nernst transport in a rutile-inspired altermagnet~\cite{hoyer2025}, magnon orbital moments and the orbital Nernst effect in a honeycomb antiferromagnet~\cite{goorbital2024}, and topological magnon--phonon hybridization in square-lattice and Fe$_2$Mo$_3$O$_8$ models~\cite{go,bao}.

\subsection{Magnon topology, boundary states, and thermal Hall response}
\label{sec:boundary}
Dzyaloshinskii--Moriya interaction (DMI) opens a topological magnon gap in the honeycomb ferromagnet and supports chiral edge modes. We use the model of Pantale\'on and Xian~\cite{pantaleon} to reproduce the published zigzag boundary spectrum and calculate the bulk Berry curvature, Chern numbers, and thermal Hall response. The spin Hamiltonian is
\begin{equation}
 H_{\mathrm F}=-J\sum_{\langle ij\rangle}\bm S_i\cdot\bm S_j
 +D\sum_{\langle\!\langle ij\rangle\!\rangle}
 \nu_{ij}\hat{\bm z}\cdot(\bm S_i\times\bm S_j).
 \label{eq:model_ferromagnet}
\end{equation}
Here $J>0$ is the nearest-neighbor ferromagnetic exchange, and $D$ is the next-nearest-neighbor DMI. Each bond is counted once, with $\nu_{ji}=-\nu_{ij}$. For the oriented next-neighbor vectors $(\bm a_1,-\bm a_2,\bm a_2-\bm a_1)$, the input uses $\nu_{ij}=+1$ on sublattice A and $-1$ on B. The vectors $\bm a_1,\bm a_2$ are primitive lattice vectors. The spins have $S=1$ and align along $+z$. Energies are measured in $JS$, and the nearest-neighbor distance is $a=1$. The zigzag strips use the published parameters $D/J=0$ and $0.1$ and the outermost-site magnon potential $\delta JS$ with $\delta=1$.

\begin{figure}[!htbp]
\centering\figfile{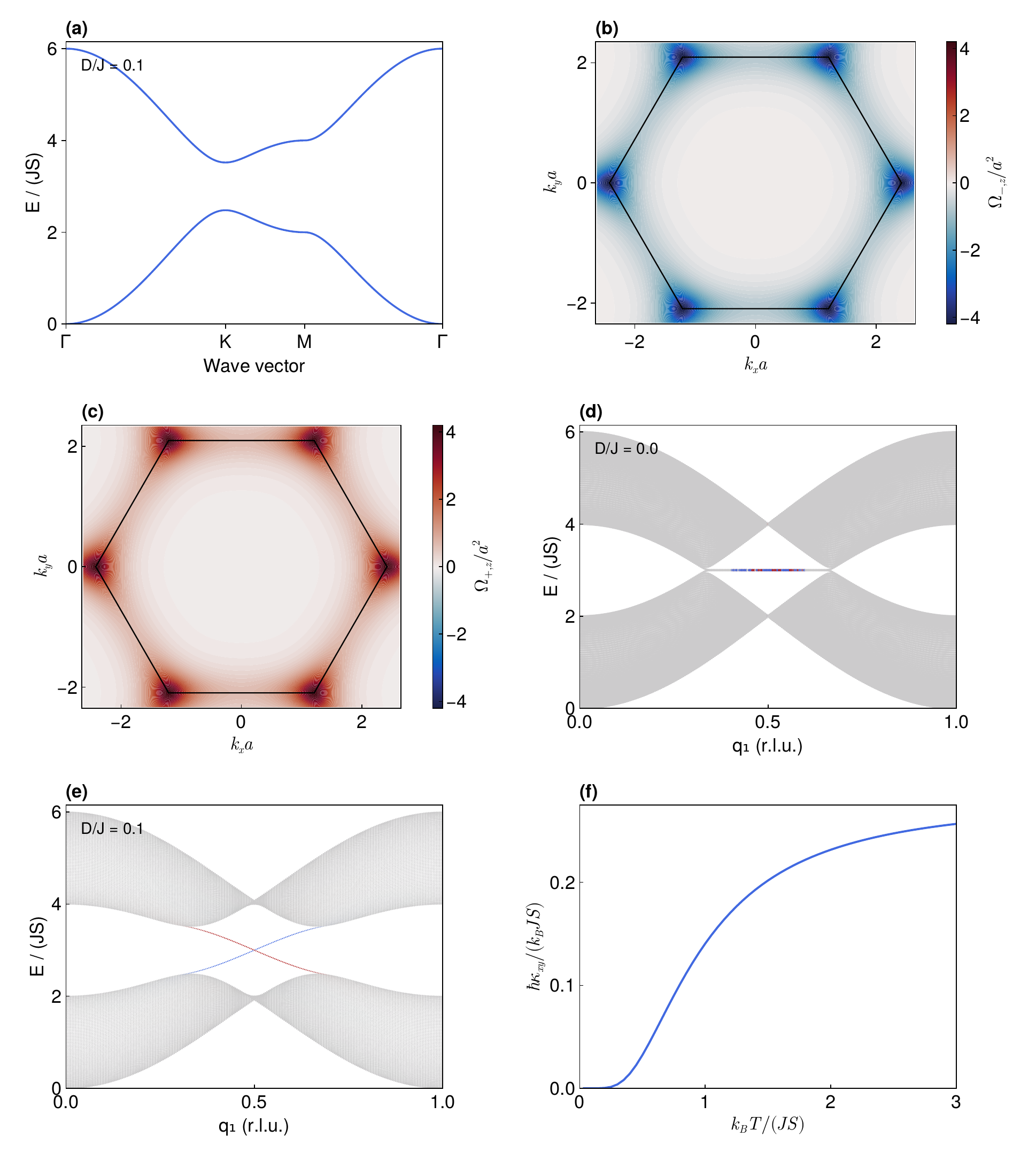}{0.98}
\caption{Magnon topology and thermal transport. (a) Bulk bands for $D/J=0.1$, with Chern numbers $(-1,+1)$ in ascending energy order. (b,c) Berry curvature of the lower and upper bands, with the first Brillouin zone outlined. Here $a$ is the nearest-neighbor distance. (d,e) Finite-strip spectra reproducing the $D=0$ and $D/J=0.1$ cases of Fig.~2(a,b) in Ref.~\cite{pantaleon}. The zigzag strips have 64 and 96 layers and the published edge potential $\delta=1$. Red and blue distinguish opposite edges, and gray denotes bulk-like modes. Here $q_1=\sqrt3 k/(2\pi)$. (f) Thermal Hall coefficient of the same bulk model at $D/J=0.1$.}
\label{fig:magnon_topology}
\end{figure}

At $D/J=0.1$, the lower and upper bulk bands carry Chern numbers $(-1,+1)$ in the convention of Eq.~\eqref{eq:chern} [Fig.~\ref{fig:magnon_topology}(a)]. Their Berry curvatures have opposite signs and concentrate near the corners of the hexagonal Brillouin zone [panels~(b,c)]. The bulk spectrum retains its acoustic Goldstone point, while boundary modes appear in the gap between the two positive-frequency branches.

Without DMI, the gapless system has an approximately flat edge branch. Finite DMI opens a bulk gap traversed by branches localized on opposite edges. Panels~(d,e) show zigzag strips with 64 layers for $D=0$ and 96 layers for $D/J=0.1$, using the termination of Ref.~\cite{pantaleon}. The localized branches cross inside the bulk gap at $q_1=1/2$, $E=3JS$.

Thermal occupation weights the opposite Berry curvatures of the two bands unequally, producing the thermal Hall response in panel~(f). The coefficient increases from zero over the plotted temperature range. With $\Theta=\kb T/(JS)$ and $\widetilde\kappa_{xy}=\hbar\kappa_{xy}/(\kb JS)$, we obtain $\widetilde\kappa_{xy}\simeq0.141$ at $\Theta=1$.

\subsection{Nonlinear magnon thermal Hall response}
\label{sec:nonlinear_magnon}
A magnetic field lifts the degeneracy of opposite-spin magnons in a honeycomb antiferromagnet and can activate a transverse heat current quadratic in the temperature gradient. We examine the extended Berry curvature dipole mechanism proposed by Ni et~al.~\cite{ni2025}, using a realization with unequal nearest-neighbor exchange couplings. \pkg\ calculates the dipole texture and the temperature and field dependence of the response. The spin Hamiltonian is
\begin{equation}
 \begin{aligned}
 H_{\mathrm{AF}}={}&\sum_{\langle ij\rangle}J_{ij}\bm S_i\cdot\bm S_j
 +D_z\sum_{\langle\!\langle ij\rangle\!\rangle}
 \nu_{ij}\hat{\bm z}\cdot(\bm S_i\times\bm S_j)\\
 &+K\sum_i(S_i^z)^2-\mathcal B\sum_iS_i^z.
 \end{aligned}
 \label{eq:model_antiferromagnet}
\end{equation}
The sums count each bond once. For this model's next-neighbor vectors $(\bm a_1,-\bm a_2,\bm a_2-\bm a_1)$, $\nu_{ij}=-1$ on A and $+1$ on B, with the opposite sign for reversed bonds. The two sublattices carry $S=1$ moments along opposite $z$ directions. We use $D_z=-0.47$~meV, easy-axis anisotropy $K=-0.45$~meV, and nearest-neighbor exchanges $(12,12,10.8)$~meV, assigning the third value to the bond along $x$. This bond anisotropy breaks threefold rotation. The Zeeman parameter $\mathcal B$ is reported in energy units.

The local extended Berry curvature dipole is $\overline D_{n,y}^{\,z}=\partial_{k_y}(E_n\Omega_n^z)$. Its energy-weighted form enters Eq.~\eqref{eq:nonlinear_thc}, which defines $j_x^{E,(2)}=\chi_{xyy}(\partial_yT)^2$. The response combines the calculated curvature and its momentum derivative with the Bose thermal weight. We use the triangular Bravais lattice length $a=1$~\AA\ and relaxation time $\tau=1$~ps, giving a two-dimensional sheet coefficient in $\mathrm{W\,m\,K^{-2}}$.

\begin{figure}[!htbp]
\centering\figfile{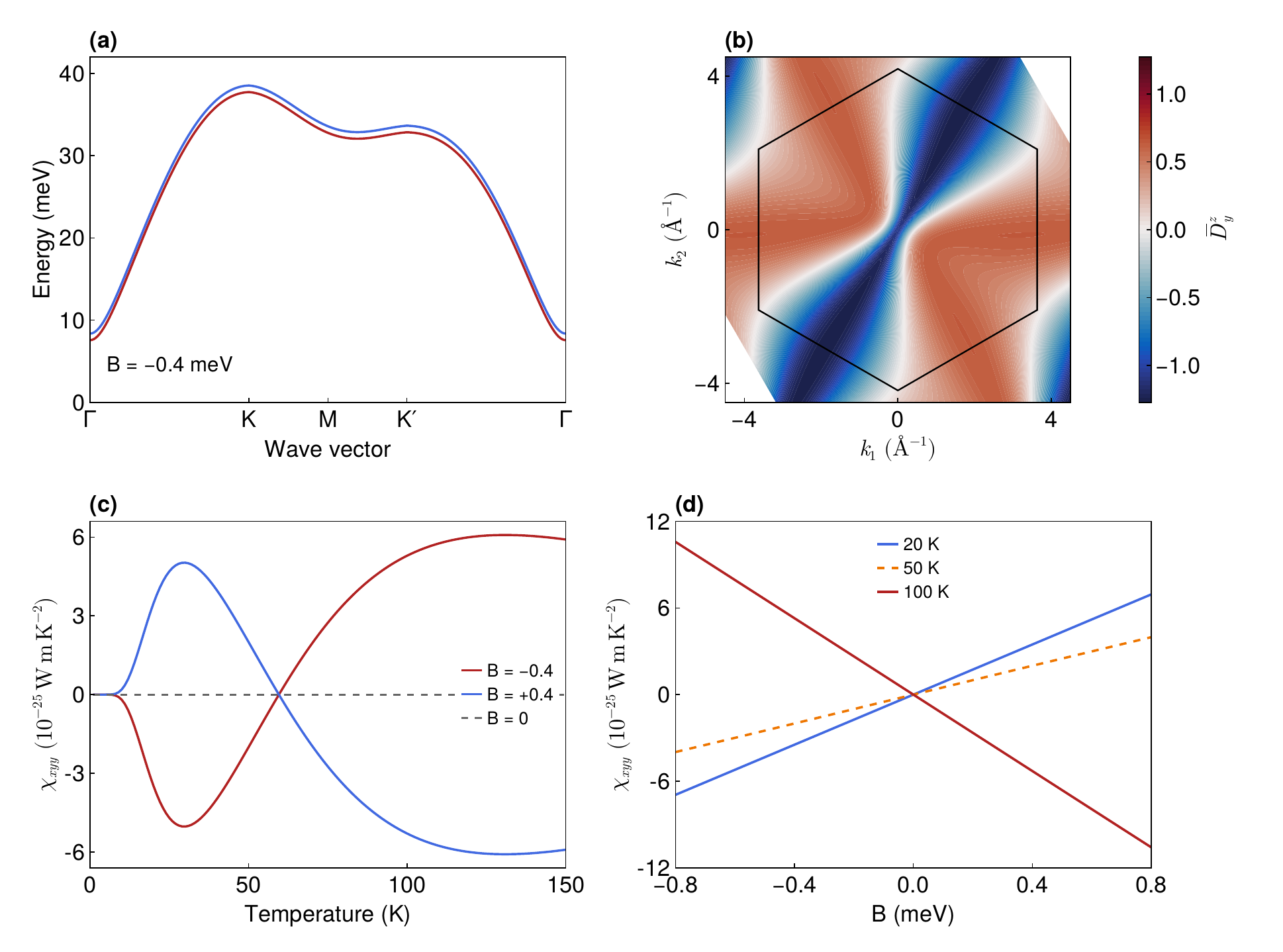}{0.98}
\caption{Nonlinear magnon thermal Hall response. (a) Bands at $\mathcal B=-0.4$~meV. (b) Lower-band extended dipole at the same field, in $\mathrm{meV}\,\text{\AA}^{3}$, with the first Brillouin zone outlined in black. (c) Temperature dependence at $\mathcal B=-0.4$, $0$, and $+0.4$~meV for the same exchange parameters. (d) Field dependence at 20, 50, and 100~K. The field variable is the Zeeman energy $\mathcal B$ in meV. Responses use $\tau=1$~ps and $120^2$ primitive-cell grids.}
\label{fig:nonlinear_magnon}
\end{figure}

At $\mathcal B=-0.4$~meV, the field separates the otherwise degenerate conserved-spin sectors by $0.8$~meV [Fig.~\ref{fig:nonlinear_magnon}(a)]. Panel~(b) shows the lower-band extended dipole over the first Brillouin zone. The plot axes $k_1,k_2$ are orthonormal coordinates in the reciprocal plane, and the derivative is taken along Cartesian $y$. The response integral covers one reciprocal primitive cell.

At zero field, the contributions from the two spin sectors cancel. Reversing the field changes the sign of $\chi_{xyy}$. Panel~(c) shows the temperature dependence from 1 to 150~K at $\mathcal B=-0.4$, $0$, and $+0.4$~meV with the same exchange parameters. At $\mathcal B=-0.4$~meV and 100~K, a $120^2$ midpoint grid gives $\chi_{xyy}\simeq5.29\times10^{-25}\,\mathrm{W\,m\,K^{-2}}$. The sign change between 59 and 60~K reflects the changing thermal weights of contributions across momentum space.

The field scans in panel~(d) span $-0.8\leq\mathcal B\leq0.8$~meV at 20, 50, and 100~K. They use field steps of $0.1$~meV and the same $120^2$ grids.

\subsection{Phonon Berry curvature and angular momentum Hall response}
\label{sec:phonon_geometry}
Two elastic models illustrate phonon geometry and angular momentum transport. The honeycomb model exhibits valley angular momentum and Berry curvature, and the simple-cubic model gives the angular momentum Hall response. We calculate both with \pkg\ using the harmonic bond Hamiltonian
\begin{equation}
 \begin{aligned}
 H_{\mathrm{ph}}={}&\sum_i\frac{\bm p_i^2}{2M_i}
 +\frac12\sum_{(ij)\in\mathcal E}\Big[
 \kappa_{\mathrm T,ij}|\Delta\bm u_{ij}|^2\\
 &\qquad +(\kappa_{\mathrm L,ij}-\kappa_{\mathrm T,ij})
 (\hat{\bm e}_{ij}\cdot\Delta\bm u_{ij})^2\Big],
 \end{aligned}
 \label{eq:model_springs}
\end{equation}
where $\Delta\bm u_{ij}=\bm u_i-\bm u_j$, $\hat{\bm e}_{ij}$ is the unit vector along the equilibrium bond, and $\mathcal E$ contains each undirected bond once. The longitudinal and transverse force constants are $\kappa_{\mathrm L,ij}$ and $\kappa_{\mathrm T,ij}$. We denote their numerical inputs by $K_{\mathrm L}$ and $K_{\mathrm T}$ after conversion to the model units through $D^{(E)}=\hbar^2D^{(\omega)}$.

For the honeycomb AB model of Zhang and Niu~\cite{zhangniu}, $\mathcal E$ contains the nearest-neighbor A--B bonds and $\bm u_i=(u_i^x,u_i^y)$. The primitive lattice constant is $a$. Mass is measured in units of $M_A$, with $M_B/M_A=1.2$, and the numerical spring parameters are $K_{\mathrm L}=1$, $K_{\mathrm T}=0.25$. The energy unit is $E_0=\hbar\sqrt{\kappa_{\mathrm L}/M_A}$.

\begin{figure}[!htbp]
\centering\figfile{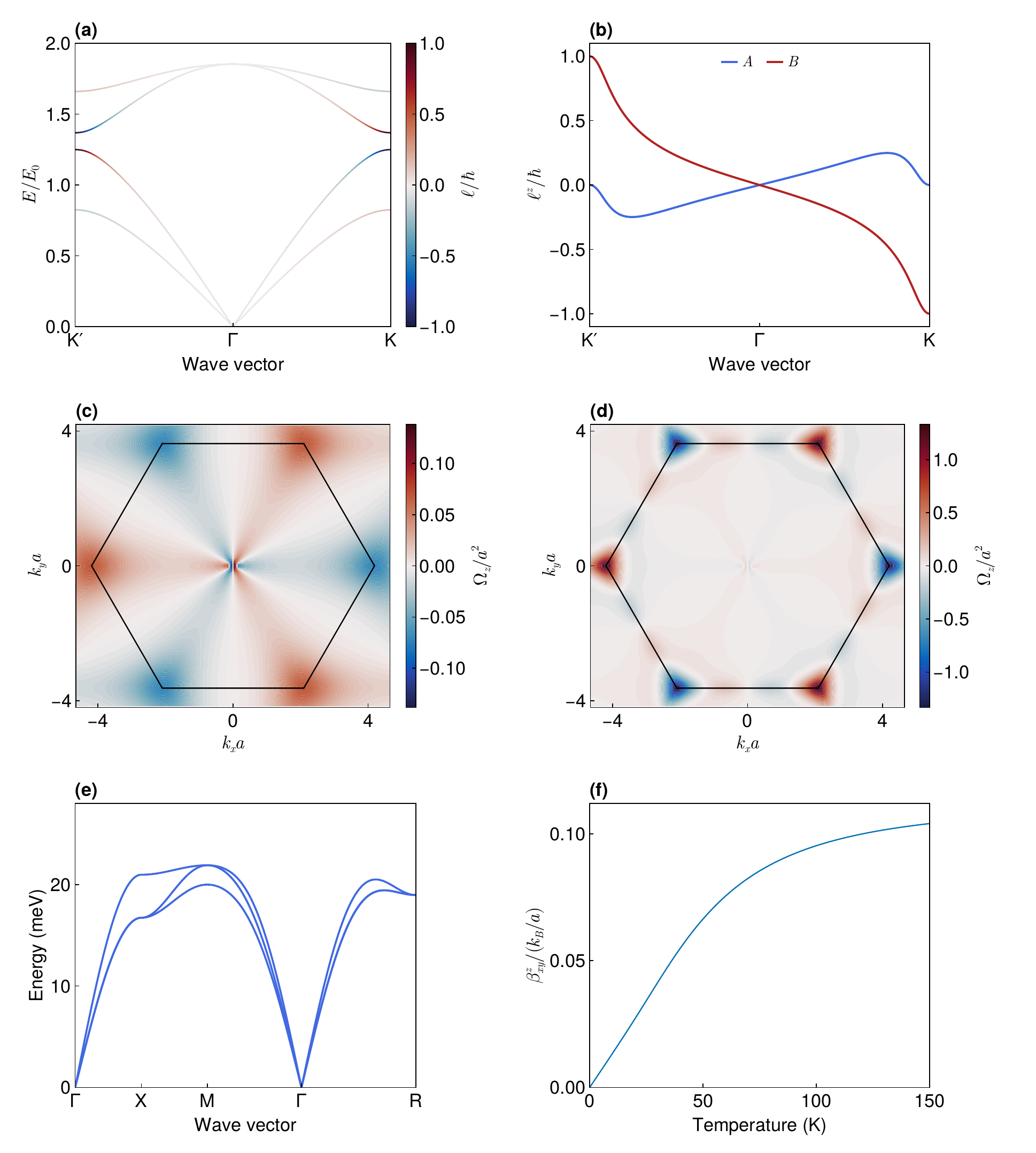}{0.99}
\caption{Phonon geometry and angular momentum Hall response. (a) Honeycomb AB bands~\cite{zhangniu} colored by mechanical PAM. (b) Atomic contributions to the second branch along the same path. (c,d) Ordinary Berry curvature of bands 1 and 2, with separate color scales and the first Brillouin zone outlined in black. (e) Bands of the simple-cubic model~\cite{parkyang}. (f) Intrinsic PAM Hall coefficient of the simple-cubic model over $0$--$150$~K, with the analytic zero-temperature limit included. The response in (f) is calculated from the angular momentum current curvature.}
\label{fig:phonon_geometry_hall}
\end{figure}

The two valleys carry opposite phonon angular momenta [Fig.~\ref{fig:phonon_geometry_hall}(a,b)]. At $K$, the second branch has $(\ell_A^z,\ell_B^z)=(0,-1)\hbar$, so its circular motion is carried by the B atom. The atomic contributions sum to the total mechanical PAM. The Berry curvatures of the two lowest branches, evaluated on a $96^2$ grid over the first Brillouin zone, also change sign between opposite valleys [panels~(c,d)], as expected from time-reversal symmetry.

The simple-cubic model of Park and Yang~\cite{parkyang} uses Eq.~\eqref{eq:model_springs} with three displacement components per site. The set $\mathcal E$ contains the six nearest neighbors and twelve face-diagonal next-nearest neighbors of each site, with each bond counted once in the energy. For unit mass, the effective spring parameters entering $D^{(E)}$ are $(K_{\mathrm L},K_{\mathrm T})=(50,20)\,\mathrm{meV}^2$ for nearest neighbors and $(20,10)\,\mathrm{meV}^2$ for next-nearest neighbors. The onsite force constants enforce the acoustic sum rule. Off-diagonal angular momentum matrix elements generate a finite Hall response through Eq.~\eqref{eq:beta}, while the equilibrium diagonal PAM vanishes. Figure~\ref{fig:phonon_geometry_hall}(e,f) shows the spectrum and Hall response, respectively.

The angular momentum Hall coefficient increases with temperature [panel~(f)], reaching approximately $\PAMHundred\,\kb/a$ at 100~K and $\PAMHundredFifty\,\kb/a$ at 150~K. The corresponding values read from Fig.~2(b) of Ref.~\cite{parkyang} are approximately $0.0943\,\kb/a$ and $0.103\,\kb/a$, respectively. We calculate the curve from 1 to 150~K with \pkg's native adaptive integration and cubic symmetry averaging, starting from a $\PAMMesh^3$ midpoint grid.

\subsection{Phonon angular momentum transport in altermagnets}
\label{sec:hybrid_boltzmann}
Magnon--phonon coupling in a $d$-wave altermagnet can transfer the alternating magnon spin texture to the mechanical phonon angular momentum of the hybrid modes. Bendin et~al.~\cite{bendin} calculated this texture in a square-lattice model and predicted a phonon angular momentum splitter effect from symmetry. That study did not evaluate the corresponding transport coefficients. We use \pkg\ to calculate the Boltzmann response tensor and its temperature dependence. The tensor also gives the longitudinal and transverse currents as the temperature gradient rotates.

Using the notation of Ref.~\cite{bendin}, the coupled model is $\hat H=\hat H_{\mathrm m}+\hat H_{\mathrm{ph}}+\hat H_{\mathrm{mpc}}$. Its spin Hamiltonian, given in Eq.~(A1) of that reference, is
\begin{equation}
 \begin{aligned}
 \hat H_{\mathrm m}={}&-\frac{J_{\mathrm{AFM}}}{2}
 \sum_{\langle i,j\rangle}\bm S_i\cdot\bm S_j\\
 &-\sum_{l=1,2}\frac{J_{\mathrm{FM},l}}{2}
 \sum_{\langle\!\langle i,j\rangle\!\rangle_l}\bm S_i\cdot\bm S_j.
 \end{aligned}
 \label{eq:hybrid_spin_model}
\end{equation}
Here $J_{\mathrm{AFM}}<0$ couples nearest neighbors antiferromagnetically, while $J_{\mathrm{FM},1}$ and $J_{\mathrm{FM},2}>0$ describe the two directional next-neighbor bond families, interchanged between the opposite-spin sublattices. The bond sums follow the convention of the reference.

The phonon Hamiltonian follows Eq.~(B1) of the same reference,
\begin{equation}
 \begin{aligned}
 \hat H_{\mathrm{ph}}={}&\sum_i\frac{\bm p_i^2}{2M}
 +\frac12\sum_{\langle i,j\rangle}\Big[
 (K_{\mathrm L}^{(1)}-K_{\mathrm T})
 (\bm u_{ij}\cdot\hat{\bm R}_{ij})^2+K_{\mathrm T}\bm u_{ij}^2\Big]\\
 &+\frac{K_{\mathrm L}^{(2)}}2
 \sum_{\langle\!\langle i,j\rangle\!\rangle}
 (\bm u_{ij}\cdot\hat{\bm R}_{ij})^2,
 \end{aligned}
 \label{eq:hybrid_phonon_model}
\end{equation}
where $\bm u_{ij}=\bm u_j-\bm u_i$, $\hat{\bm R}_{ij}$ is the unit bond vector, and the displacements and momenta lie in the $xy$ plane. The displacement-induced DMI coupling is the reference's Eq.~(2),
\begin{equation}
 \hat H_{\mathrm{mpc}}=\frac{DS}{R}
 \sum_{\langle\!\langle i,j\rangle\!\rangle}
 \left\{\pm(\bm u_i-\bm u_j)\cdot(\bm S_i-\bm S_j)\right\}.
 \label{eq:hybrid_coupling_model}
\end{equation}
The coupling runs over same-sublattice next-neighbor bonds of length $R=\sqrt2a$, with $+$ for the spin-up sublattice and $-$ for the spin-down sublattice. The full quadratic hybrid Hamiltonian retains anomalous terms.

The dimensionless model parameters are $S=1$, $M=10$, $J_{\mathrm{AFM}}=-1$, $J_{\mathrm{FM},1}=0.383$, $J_{\mathrm{FM},2}=0.5$, $K_{\mathrm L}^{(1)}=160$, $K_{\mathrm T}=K_{\mathrm L}^{(2)}=40$, and $D=0.25$, with $E_0=|J_{\mathrm{AFM}}|$ and $a=1$. The ferromagnetic exchange coefficients used in Eq.~\eqref{eq:hybrid_spin_model} are $2J_{\mathrm{FM},l}$, matching the published uncoupled magnon dispersion. They are twice the coefficients printed in Eq.~(S.4b) of the source supplement.

\begin{figure}[!htbp]
\centering\figfile{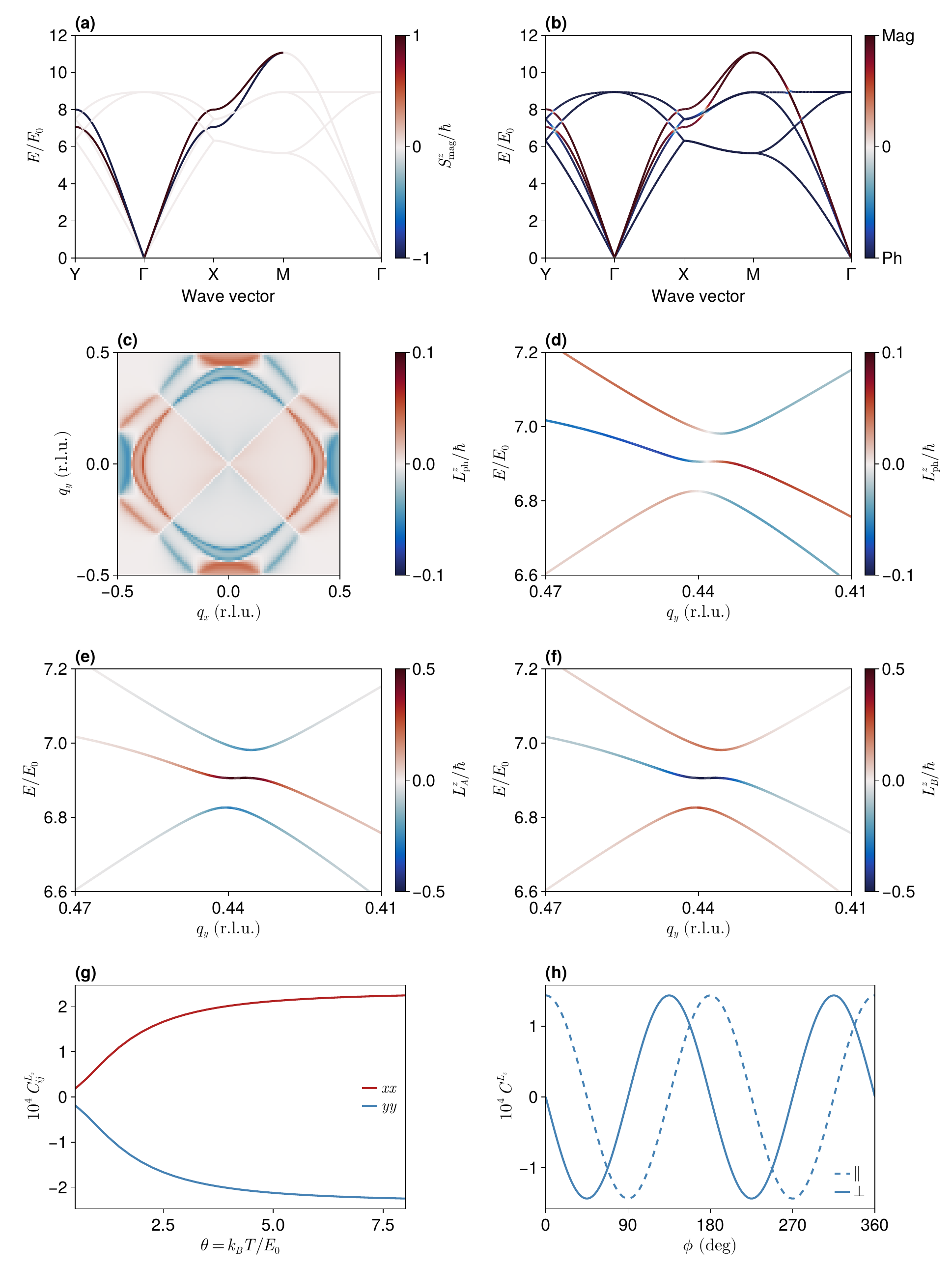}{0.90}
\caption{Phonon angular momentum currents in the square-lattice hybrid model. (a) Uncoupled modes colored by magnetic spin. (b) Six hybrid branches colored by mode character. (c) Mechanical PAM of band~4 over the magnetic Brillouin zone. (d--f) The anticrossing along $q_x=0$, colored by the total PAM and its sublattice A and B contributions. The total and sublattice color ranges are $\pm0.1\hbar$ and $\pm0.5\hbar$, respectively. (g) Boltzmann coefficients obtained from the conventional symmetrized current, per common relaxation time. (h) Longitudinal and transverse coefficients at $\theta=2$, with directions defined in the text. The transport curves use a $256^2$ grid.}
\label{fig:hybrid_boltzmann}
\end{figure}

Hybridization mixes two magnetic modes with four in-plane phonon modes to form six positive-frequency branches [Fig.~\ref{fig:hybrid_boltzmann}(a,b)]. Band~4, numbered in ascending energy order, acquires the $d$-wave mechanical PAM texture shown in panel~(c). Across the anticrossing in panels~(d--f), the decomposition $L^z_{\mathrm{ph}}=L_A^z+L_B^z$ shows the partial cancellation of the two sublattice contributions.

The \texttt{boltzmann\_transport} interface uses the conventional symmetrized current $\hat J_i^{L_z}=\{\hat L_{\mathrm{ph}}^z,\hat v_i\}/2$ to evaluate Eq.~\eqref{eq:boltzmann}. The sum includes all six bands and retains the full normalization of each hybrid mode. We report the response per common relaxation time. With $\theta=\kb T/E_0$ and $\bar\tau=\tau E_0/\hbar$, the coefficient $C_{ij}^{L_z}=\beta_{ij}^{L_z}/(\kb\bar\tau)$ gives
\begin{equation}
 \frac{j_i^{L_z}}{E_0/a}
 =-\bar\tau C_{ij}^{L_z}\frac{\partial\theta}{\partial(x_j/a)}.
 \label{eq:hybrid_boltzmann_units}
\end{equation}
The Boltzmann tensor is diagonal and satisfies $C_{yy}^{L_z}=-C_{xx}^{L_z}$, producing opposite angular momentum responses along the two crystal axes [panel~(g)]. At $\theta=2$, $C_{xx}^{L_z}\simeq1.44\times10^{-4}$ for the symmetrized current. The semiclassical current $L_n^zv_{n,i}$ gives $C_{xx}^{L_z}\simeq3.34\times10^{-4}$. The calculation covers the full magnetic Brillouin zone with a $256^2$ midpoint grid and 31 temperatures from $\theta=0.5$ to~8.

The opposite diagonal tensor entries determine the angular momentum splitter response in panel~(h), shown at $\theta=2$. For a temperature gradient along $\bm e=(\cos\phi,\sin\phi)$ and transverse direction $\bm p=(-\sin\phi,\cos\phi)$, the projections are $C_\parallel=\bm e^TC\bm e=A\cos2\phi$ and $C_\perp=\bm p^TC\bm e=-A\sin2\phi$, with $A=C_{xx}^{L_z}$. Both current components are proportional to the common relaxation time.

\FloatBarrier
\section{Conclusions}
\pkg\ calculates spectra, topology, and transport for magnons, phonons, and their hybrid modes with a consistent bosonic normalization. Native phonon and spin--lattice modules complement Sunny spin-wave models. The examples reproduce published magnon boundary spectra and show how a magnetic field controls the nonlinear magnon thermal Hall response. They also illustrate phonon band geometry and angular momentum Hall transport.

For the altermagnetic model, we calculate transport coefficients associated with the published $d$-wave phonon angular momentum texture. The original study~\cite{bendin} predicted the corresponding splitter effect from symmetry but did not evaluate these coefficients. Within the relaxation-time approximation, the native Boltzmann interface gives the temperature-dependent response tensor and its longitudinal and transverse projections.

\section*{Acknowledgements}
This work was supported by the National Natural Science Foundation of China (Grant No.~52130204).
\section*{CRediT authorship contribution statement}
\textbf{Xiaonong Shen:} Software, Validation, Visualization, Writing -- original draft.\par
\textbf{Wei Ren:} Supervision, Funding acquisition, Writing -- review \& editing.
\section*{Declaration of competing interest}
The authors declare no competing interests.
\section*{Data and code availability}
The source code of \pkg\ v0.2.4 and the model inputs used to reproduce the figures are provided in the accompanying software archive.
\appendix
\section{Example input: phonon angular momentum Hall response}
\label{app:example_input}
The following input constructs the simple-cubic model in Sec.~\ref{sec:phonon_geometry} and plots its bands. It also plots the angular momentum current curvature on the $q_z=0$ plane. Adaptive three-dimensional integration then gives the Hall response as a function of temperature.

\begin{lstlisting}
using Sunny, MagBerry, CairoMakie
crystal = Crystal(lattice_vectors(1,1,1,90,90,90),
                  [[0,0,0]], 221)
phonon = PhononSystem(crystal, [1.0])
set_spring!(phonon, 50, Bond(1,1,[1,0,0]); transverse=20)
set_spring!(phonon, 20, Bond(1,1,[1,1,0]); transverse=10)
enforce_asr!(phonon)

path = q_space_path(crystal,
    [[0,0,0],[.5,0,0],[.5,.5,0],[0,0,0],[.5,.5,.5]],301;
    labels=["(*@$\Gamma$@*)","X","M","(*@$\Gamma$@*)","R"])
plot_bands(phonon, path)

bc = berry_curvature(phonon; kmesh=(40,40), observable=:pam)
plot_berry_curvature(bc; band=3)

plot_angular_momentum_nernst_conductivity(phonon;
    kmesh=10, temperatures=1:150, adaptive_refinement=true)
\end{lstlisting}

\section*{Declaration of generative AI use}
OpenAI Codex was used to assist drafting and language revision of this manuscript. The authors reviewed the resulting text, made the necessary revisions, and are responsible for the final content.

\end{document}